# A time grating approach to ultrahigh-Q guided mode resonance

Youxiu Yu[1], Xiaofeng Xu[2], Yang Long[3], Gui-Geng Liu[4], Dongliang Gao[2], Xiao Lin[5], and Hao Hu[1,*]

[1]*National Key Laboratory of Microwave Photonics & College of Electronic and Information Engineering, Nanjing University of Aeronautics and Astronautics, Nanjing 211106, China*

[2]*College of Physical Science and Technology, Soochow University, Suzhou 215006, China*

[3]*School of Physics Science and Engineering, Tongji University, Shanghai 200092, China*

[4]*Department of Electronic and Information Engineering, School of Engineering, Westlake University, Hangzhou 310030, China*

[5]*Interdisciplinary Center for Quantum Information, State Key Laboratory of Modern Optical Instrumentation, Zhejiang University, Hangzhou 310027, China*

*Corresponding author:* hao.hu@nuaa.edu.cn

Guided mode resonance (GMR), the resonant coupling of free-space light into leaky waveguide modes, is traditionally achieved with periodic patterned structures. However, this approach makes its key properties such as quality factor (Q-factor) fabrication-dependent and non-tunable. Here, we introduce a time grating platform, i.e., a homogeneous waveguide whose refractive index is modulated periodically in time, that allows tunable GMRs through temporal modulation engineering rather than spatial structural redesign. We show that the Q-factors of these GMRs diverge as the modulation depth vanishes. Furthermore, unconstrained by energy conservation, the resonances exhibit near-unity reflection for fundamental harmonics and values exceeding 40 for first-order harmonics. Our findings not only apply to yield a giant Goos-Hänchen shift over $10^3$ times wavelength without sacrificing the reflection magnitude, but also open new avenues for related phenomena such as bound states in the continuum, unidirectional GMRs and beyond.

# 1. Introduction

Guided mode resonance (GMR) occurs when free-space waves couple into guided modes through periodic structures such as gratings or photonic crystal slabs.[1,2] Owing to its direct accessibility from free space, GMR has enabled a broad range of applications, including molecular sensing, spectral filtering, light emission, etc.[1,3–5] However, the inherently leaky nature of GMR inevitably leads to radiation losses, fundamentally constraining the achievable Q-factor and compromising the performance of GMR-based devices. These losses become even more pronounced in practical implementations, where fabrication imperfections such as structural defects and disorders further exacerbate the mode leakage. Consequently, mitigating radiation losses remains a key challenge in the design of high-Q GMR-based photonic devices.

Subsequent studies have demonstrated that radiation losses of GMRs can be entirely suppressed at specific wavevectors, giving rise to a class of states known as bound states in the continuum (BICs).[6–8] These states feature theoretically infinite Q-factors due to their complete decoupling from free-space radiation. However, such a perfect confinement also precludes any energy exchange between the guided modes and external waves. To achieve controlled radiation while retaining high Q-factors, a moderate periodic perturbation is typically introduced into the BIC system. Unfortunately, as previously mentioned, fabrication imperfections in these patterned structures would induce unwanted radiation losses, greatly limiting the stability of the Q-factors. Thus, it is interesting to ask whether there is an alternate mechanism to enhance the robustness and environmental tolerance of high-Q GMR systems.

Recently, time-varying media (such as time photonic crystals and time gratings) have been recognized as transformative platforms for wave manipulation.[9–13] Unlike their spatial counterparts, these media feature constitutive parameters (e.g., permittivity and conductivity) that are rapidly modulated in time through optical or electrical signals.[14,15] Researches reveal that time-varying media not only replicate a variety of electromagnetic phenomena observed in static systems (e.g., double-slit diffraction, Anderson localization, topological bulk-edge correspondence),[16–19] but also unlock unconventional behaviors such as broadband frequency translation, coherent wave control with different incident frequencies, non-resonant amplification, and threshold-free Cherenkov radiation.[20–24] Crucially, because they do not rely on spatially patterned structures, time-varying systems are inherently robust against fabrication imperfections. Their high degree of tunability and temporal reconfigurability offers enhanced environmental adaptability, making them promising candidates for next-generation photonic devices. This naturally raises the question: can time-varying media be harnessed to achieve GMR with enhanced Q-factor control and resilience?

In this work, we propose a time grating approach to achieve ultrahigh-Q GMR. The time grating, i.e., a homogeneous dielectric waveguide with permittivity periodically modulated in time, supports GMR when the energy mismatch between free-space waves and guided modes is compensated by the modulation frequency. Remarkably, the Q-factor of the resulting GMR can be readily controlled by adjusting the modulation depth. For instance, as the modulation depth is reduced to 0.005, we observe an ultrahigh Q-factor exceeding $10^5$. Moreover, because the temporal modulation breaks the energy conservation, the system exhibits highly nontrivial reflection behaviors: near-unity reflection coefficient is achieved for the fundamental harmonic, while the first-order harmonic reflection coefficient can even exceed 40. The combination of ultrahigh Q-factors and amplified reflection makes time-grating-based GMR an indispensable ingredient for Goos-Hänchen (GH) shift enhancement, overcoming the traditional trade-off between lateral shift magnitude and reflection. These findings open new avenues for dynamically reconfigurable, fabrication-tolerant photonic devices and pave the way toward time-modulated platforms for advanced sensing, signal processing, and active wavefront control.

## 2. Principle

Without loss of generality, we consider a time grating made of a homogeneous dielectric waveguide with a sinusoidal modulation of permittivity in time. The permittivity of dielectric waveguide is expressed as $\varepsilon_r(t) = \varepsilon_{r0}[1+\delta\cos(\Omega t)]$, where $\varepsilon_{r0}$ is the unperturbed permittivity, $\delta$ is the modulation depth, and $\Omega$ is the modulation frequency. Note that the modulation frequency is used to form the time grating and has no direct relationship with the incident frequency described below. For clarity and simplicity, the term "modulation frequency" and "incidence frequency" used here specifically refer to the angular frequency. As plotted in **Figure 1**b, the surface of time grating aligns with the *y-z* plane, while the surface normal is oriented along the *x*-axis. To achieve platform-independent universality, we present our results using normalized quantities for frequencies and thicknesses. The normalized thickness of the time grating is adopted as $L_{TG} = L\sqrt{\varepsilon_{r0}}\Omega/(2\pi c)$. Unless otherwise specified, the unperturbed permittivity and modulation frequency are taken as $\varepsilon_{r0} \approx 3.2$ and $\Omega \approx 555.88$ THz, respectively, in the following analytical calculations. We emphasize that such a parameter setup is experimentally feasible by employing indium tin oxide (ITO) as the material platform, which exhibits pronounced nonlinear optical responses. First, the unperturbed permittivity of ITO is $\varepsilon_{r0} \approx 3.2$ in the wavelength band of 0.6~0.66μm (corresponding to the working frequency band of 5.1Ω~5.6Ω). Second, the rise time of ITO has been reported to be only a few femtoseconds, which allows it to instantaneously respond to the external optical pumping with the modulation frequency Ω considered here.[16] We remark that the specific parameter values adopted here are intended as representative examples rather than strict constraints of the proposed mechanism.

First, we present the excitation condition of GMRs in time grating. Initially, a free-space incidence with transverse-magnetic (TM) polarization cannot excite GMR in the waveguide without external modulations. This is straightforward because the frequency of the incident wave $\omega_i = (k_y c)/\sin\theta$ (with $\theta$ being the incident angle) is different from that of the guided mode $\omega_g$ at the same tangential wavevector $k_y$. However, by introducing a periodic temporal modulation, the system provides additional frequency compensation as $\Delta\omega = m(2\pi/T)$ to the incident wave. Here, *m* is an arbitrary integer. Such a frequency compensation couples the TM incidence to the guided mode if

$$\omega_g(k_y) = \omega_i(k_y) \pm m\frac{2\pi}{T} \tag{1}$$

where $T = 2\pi/\Omega$ corresponds to the temporal modulation period (see mode transition diagram in Figure 1d). This excitation mechanism is fundamentally different from that in a conventional space grating. In a typical space grating as shown in Figure 1a, the structural periodicity (with a period denoted as $P$) introduces an extra momentum compensation, $\Delta k = m(2\pi/P)$, to incident waves. Then, GMR is excited if the difference between the incident wavevector component $k_{y,i} = k_0 \sin\theta$ and the guided mode' one $k_{y,g}$ is compensated by $\Delta k$ at a given frequency $\omega$ (Figure 1c). In other words, the excitation condition of GMR in space grating is expressed as $k_{y,g}(\omega) = k_{y,i}(\omega) \pm m(2\pi/P)$.

## 3. Result and Discussion

The excitation of GMRs in time grating is spectrally manifested as sharp peaks in the reflection spectrum. To illustrate this, we plot in **Figure 2**b the reflection coefficient as a function of incident angle and frequency for the fundamental harmonic in time grating (For the analytical calculation of the reflection coefficient, see Supplementary Information, Section S2). Obviously, the reflection coefficient is significantly enhanced at specific incident angles and frequencies. For example, at the incident angle of 64.5°, a sharp asymmetric Fano line shape is observed at the frequency of $\omega_i = 5.3\Omega$ (see the red dot in Figure 2b and its inset). Such a Fano line shape originates from the interference between high-Q GMR and low-Q Fabry-Pérot resonance. We find that the parameters leading to the enhanced reflection are in excellent agreement with theoretical predictions using Equation (1), namely that the resonance of this structure can be simply predicted through the GMR induced by frequency compensation. For comparison, in a conventional waveguide without time modulation, the above sharp asymmetric Fano line shapes disappear in Figure 2a. These results further show that time modulation offers an alternative route for achieving GMRs.

Next, we demonstrate that the Q-factor of GMR in the time grating could be flexibly controlled by adjusting the modulation depth. The Q-factor is extracted from the reflection spectrum of the fundamental harmonic in **Figure 3**a, under an incident angle of $\theta = 64.5°$. When the modulation depth is relatively large, e.g., $\delta = 0.5$, the GMR exhibits a broad linewidth, corresponding to a small Q-factor of 22 (Figure 3(b)). By contrast, if the modulation depth is sufficiently reduced to, e.g., $\delta = 0.005$, an extremely narrow resonance linewidth emerges. Then the Q-factor is remarkably enhanced to a value exceeding $10^5$. With a further reduction of the modulation depth, the Q-factor can increase even more and theoretically approach infinity, which can be practically achieved by decreasing the intensity of the external signals. Importantly, this tunable control of Q-factors is realized without any structural reconfiguration (i.e., without requiring special spatial structures), highlighting the unique advantage of our time grating system. It noted that the time grating degenerates into a purely spatial waveguide when the modulation depth is reduced to zero. Under this condition, the Q-factor is an infinitely lager in theory due to the waveguide mode lies below the light cone.

A small modulation depth in the time grating not only results in ultrahigh-Q GMR, but also induces strong reflection at the resonant frequency and its higher-order harmonics. As the modulation depth decreases, the resonance linewidth becomes narrower; however, the maximum reflection coefficient at the fundamental harmonic consistently remains unity, as shown in Figure 3a. This phenomenon is caused by constructive interference between the GMR and the Fabry-Pérot background under time modulation, as shown in the inset of Figure 2 and Figure S1. At higher-order harmonic frequencies in the time grating, the reflection coefficient can even exceed unity. This can be learned from Figure 3c, presenting the first-order harmonic reflection coefficient as a function of incident frequency and modulation depth. Remarkably, as the modulation depth decreases, the reflection coefficients become even higher: the calculated reflection coefficients are 3.5, 4.7, 6.6, 10.2, 20.8, and 41.4 for modulation depths of 0.5, 0.4, 0.3, 0.2, 0.1, and 0.05, respectively. This unconventional reflection spectrum arises from time-modulation-induced amplification, where the energy from the modulation signal is constantly fed into the GMR (see Supplementary Information, Section S3). In sharp contrast, in a conventional grating, the reflection coefficients of high harmonics are generally small and become negligible when the structural perturbation is sufficiently weak.

Last but not least, we reveal that our proposed GMR in time grating offers an indispensable way to engineer GH shift. GH shift is known as the lateral displacement that occurs in geometric space when an incident beam impinges on the interface between two different media (see schematic in **Figure 4**a).[25,26] As mentioned in Ref. 26, GH shift is quantified as $\Delta GH = -(\lambda/2\pi \cos(\theta))d\varphi/d\theta$, where $\varphi$ is the phase of the reflection coefficient, and $\lambda$ is the incidence wavelength. Typically, the GH shift is very small (on the order of a wavelength or less). Due to this reason, while GH shift shows great promise for applications in precision measurement, optical switches, and wavelength division multiplexers,[27–29] this small magnitude makes it difficult to be detected and exploited in practical systems. Although previous work proposed Brewster effect or resonance-based mechanisms to enhance GH shift,[8,30–33] the corresponding reflection magnitude is typically too small to be detected. Therefore, achieving a balance between the GH shift length and reflection magnitude remains a fundamental challenge.

The above challenge can be effectively addressed through the GMR enabled by our time-grating platform. The GH shift as a function of incident angle and modulation depth is shown in Figure 4b. At a fixed modulation depth, the maximum GH shift is observed at the incident angle of $\theta = 64.5°$, where GMR is excited in time grating. This is because GMR induces a sharp phase variation as the incident angle varies, favorably enhancing the GH shift (see Figure 4d). Such a phase variation becomes even more drastic as the modulation depth goes to zero, and as a result, the GH shift enhancement is further optimized (Figures 4c, e). More quantitatively, the GH shift exhibits a linear dependence on the modulation depth $1/\delta^2$ (Figure 4f). Remarkably, a giant GH shift exceeds over $10^3$ times the wavelength when the modulation depth is reduced to $\delta = 0.05$ (Figure 4e). Such a giant GH shift does not sacrifice the magnitude of reflection (e.g., the reflection coefficient of the first-order harmonic is as large as 41.4 if the modulation depth is $\delta = 0.05$). These findings suggest that GMR in time grating could simultaneously enhance GH shift strength and reflection magnitude, paving the way for advanced optical manipulation and precision metrology applications.

**4. Conclusion**

In this work, we successfully extend the concept of GMR from space to time grating. Such an extension is nontrivial, as the Q-factors of GMR, initially tuned by structural reconfiguration, could now be flexibly engineered by adjusting the modulation depth in a flat waveguide. By minimizing the modulation depth, the revealed GMR not only possesses ultrahigh Q-factor up to $10^5$, but also induces strong reflection coefficients exceeding 40 for first-order harmonics. These exotic properties of GMR in time grating make it a powerful platform to enhance GH shift with strong reflection, overcoming the conventional trade-off between GH shift length and reflection magnitude. Our configuration could be implemented in a variety of physical systems. For example, the time grating can be realized in a waveguide made of nonlinear materials such as indium tin oxide (ITO), whose refractive index is varied periodically in time by employing a temporally modulated signal.[16,34] Such time grating can also be achieved in time-varying transmission lines or metasurfaces, where the refractive index is temporally controlled by adopting optoelectronic components, such as varactor diodes and photodiodes.[35–39]

Our work also inspires future exploration of rich physics related to GMR in time grating. One promising opportunity is that BIC modes, which are the limiting case of GMRs with vanishing radiation loss, may arise under proper temporal symmetry or parameter conditions. Furthermore, introducing traveling-wave modulations can break mirror symmetry and render the revealed GMR unidirectional. Such BICs, unidirectional GMRs and beyond in time gratings offer distinct advantages of strong tunability and unconventional spectral features compared to their conventional counterparts, and thus hold great potential for practical applications in the dynamic control of light-matter interactions.

**Supporting Information**

**Acknowledgements**

This work was supported by National Natural Science Foundation of China (Grant No. 12404363, 12174281), Natural Science Foundation of Jiangsu Province (Grant No. BK20241374), Distinguished Professor Fund of Jiangsu Province, and Fundamental Research Funds for the Central Universities, NUAA (Grants No. NS2024022, NE2024007).

**Conflict of Interest**

The authors declare no conflict of interest.

**Data Availability Statement**

The data that support the findings of this study are available in the supplementary material of this article.

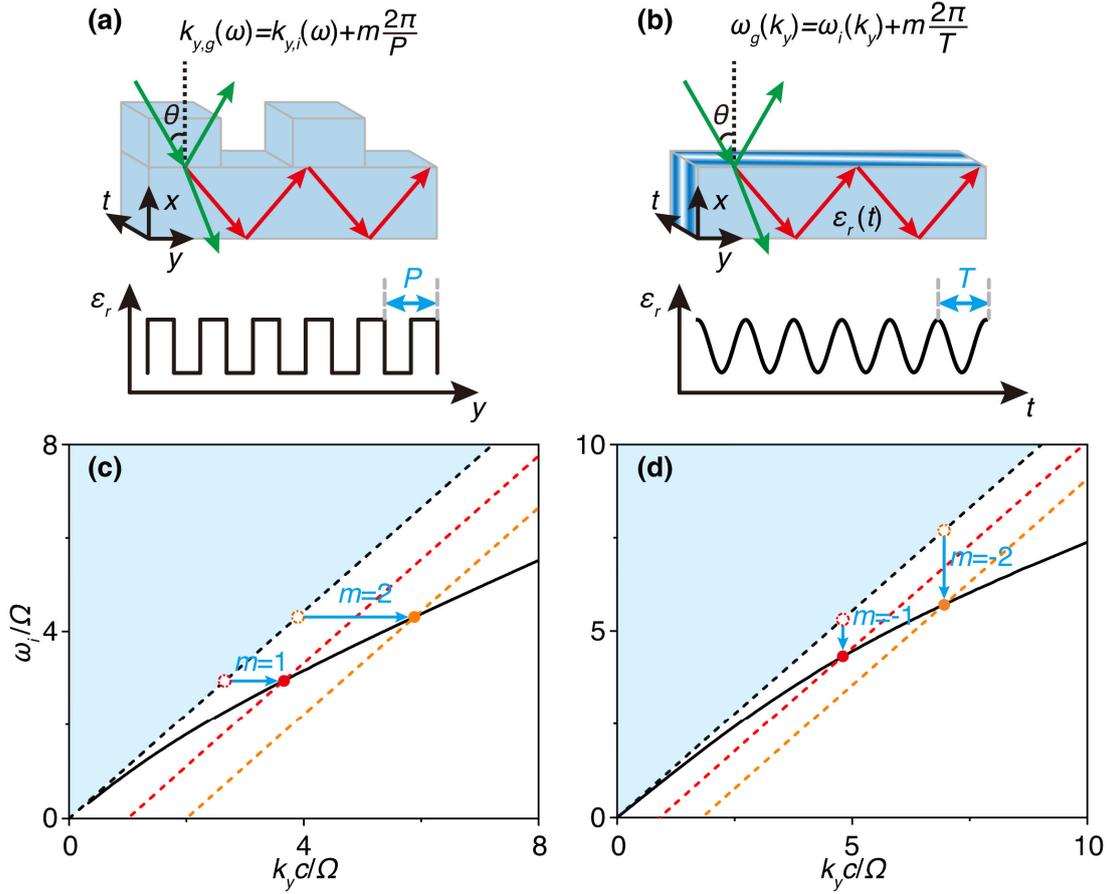

**Figure 1.** Comparison between GMRs in space grating and time grating. (a) Schematic of GMR in a space grating. (b) Schematic of GMR in a time grating. In space grating, the effective permittivity is periodically changed along $y$ axis, while in time grating, the permittivity is periodically changed along time axis. The modulation periods of space and time grating are denoted as $P = 2\pi c / \Omega$ and $T = 2\pi / \Omega$, respectively. (c) Excitation mechanism of GMR in the space grating. (d) Excitation mechanism of GMR in the time grating. In (c&d), the black solid lines depict the dispersion in waveguide, the black dashed lines depict the dispersion in air background, whereas the red and orange dashed lines indicate the dispersion relations for $m = 1, 2$ in the space grating and $m = -1, -2$ in the time grating. In all the panels, the unperturbed permittivity, modulation depth and normalized thickness of the time grating are $\varepsilon_{r0} \approx 3.2$, $\delta = 0.2$, and $L_{TG} = 1/4\pi$, respectively. The incidence angle is $\theta = 64.5°$.

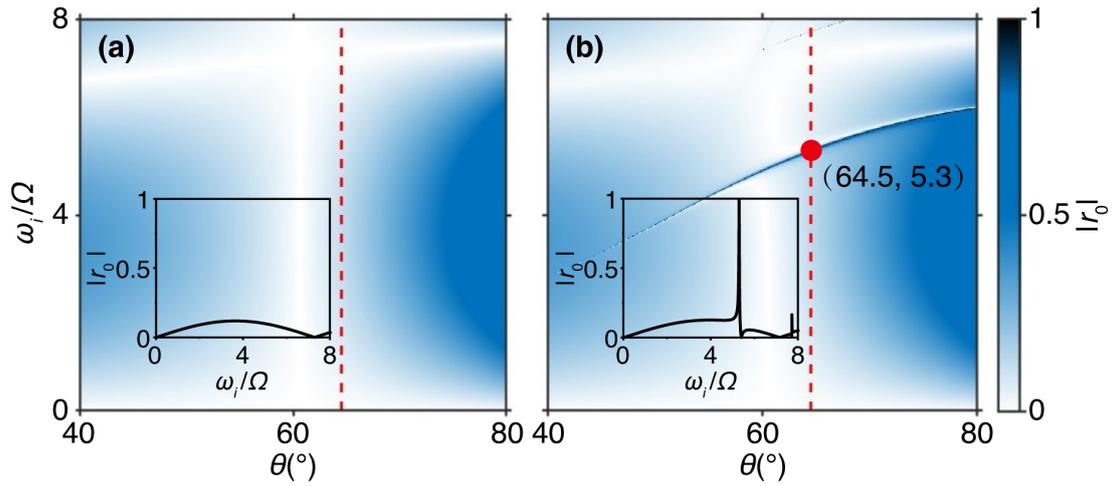

**Figure 2.** Comparison of reflection spectra in waveguides with and without time modulation. (a) Reflection spectra in waveguides without time modulation. (b) Reflection spectra in waveguides with time modulation. In (a, b), the insets correspond to the reflection coefficient as the function of the incident frequency at $\theta = 64.5°$ (as indicated by the red dashed line in (a, b)). As marked by the red dot in (b), the peak of reflection coefficient occurs at $\theta = 64.5°$ and $\omega_i = 5.3\Omega$.

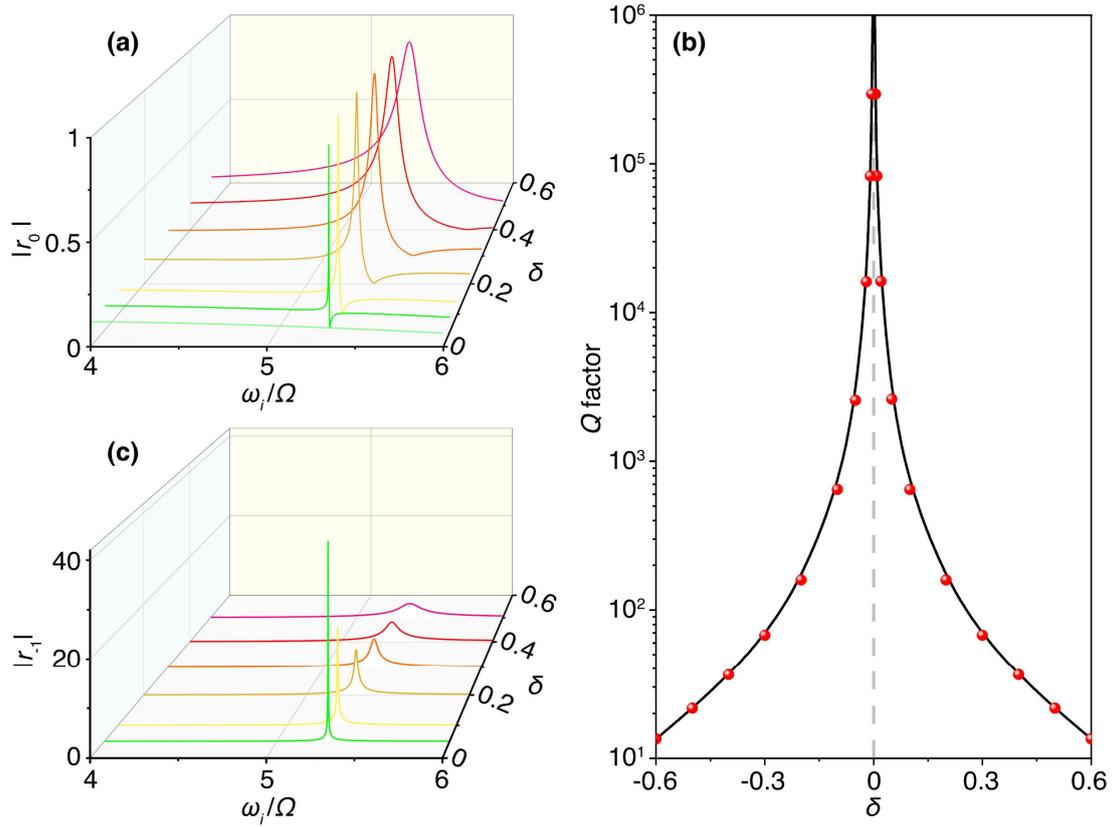

**Figure 3.** Influence of modulation depth on reflection spectra of time grating and Q-factor of GMR. (a) Influence of modulation depth on reflection spectra of time grating for fundamental harmonic. (b) Influence of modulation depth on Q-factor of GMR. (c) Influence of modulation depth on reflection spectra of time grating for first-order harmonic. In all panels, the incident angle is $\theta = 64.5°$.

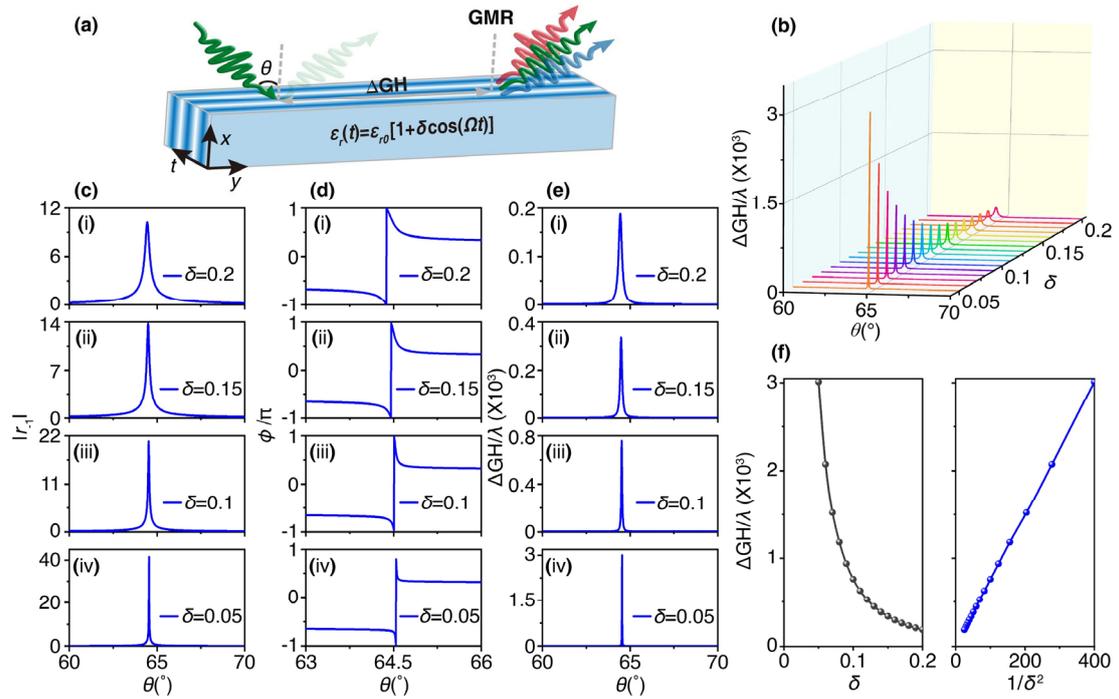

**Figure 4.** GH shift in the time grating. (a) Schematic of GH shift enhanced by GMR in the time grating. (b) GH shifts of first-order harmonic as a function of incidence angle and modulation depth in the time grating. (c) Reflection coefficient of first-order harmonic as a function of incidence angle in the time grating. (d) The phase angle of reflection coefficient as a function of incident angle in the time grating. (e) GH shift of first-order harmonic as a function of incidence angle in the time grating. (f) Relation between maximum GH shift and modulation depth. In all panels, the incident frequency is $5.3\Omega$.